\documentclass[a4paper,fleqn,usenatbib]{mnras}

\usepackage{mathptmx}
\usepackage[T1]{fontenc}
\usepackage{ae,aecompl}
\usepackage{amsmath}
\usepackage{amssymb}
\usepackage{epsfig}
\usepackage{latexsym}
\usepackage{rotating}
\usepackage{times,epsf}
\usepackage{txfonts}
\usepackage{varioref}
\usepackage{verbatim}
\usepackage{url}
\usepackage{tikz}
\usepackage{epstopdf}
\usepackage{color}
\usepackage{array}
\usepackage{multirow}
\usepackage{epstopdf}
\usepackage{float}
\usepackage{layouts}
\usepackage{url}
\usepackage{cases}
\usepackage{graphicx}
\usepackage{multicol}

\title[Levitating atmospheres of luminous neutron stars]{Levitating atmospheres of Eddington-luminosity neutron stars}

\author[M. Wielgus et al.]{
Maciek Wielgus,$^{1}$\thanks{E-mail: maciek.wielgus@gmail.com (MW)}
Aleksander S\k{a}dowski,$^{2,3}$\thanks{E-mail: asadowsk@mit.edu (AS)}
W\l{}odek Klu\'{z}niak,$^{1}$\thanks{E-mail: wlodek@camk.edu.pl (WK)}
\newauthor  \ Marek Abramowicz$^{1,4,5}$ 
and Ramesh Narayan$^{3}$
\\
$^{1}$Nicolaus Copernicus Astronomical Center,
         ul. Bartycka 18, 00-716,
         Warszawa, Poland \\
$^{2}$ MIT Kavli Institute for Astrophysics and Space Research, 77
Massachusetts Ave, Cambridge, MA 02139, USA \\
$^{3}$Harvard-Smithsonian Center for Astrophysics, 60 Garden Street,
 Cambridge, MA 02138, USA \\
 $^4$ Physics Department, Gothenburg University, 412-96 Goteborg, Sweden\\
 $^5$ Institute of Physics, Faculty of Philosophy and Science, Silesian University in Opava,  Bezrucovo nam. 13, 746-01 Opava, Czech Republic
}

\date{Accepted XXX. Received YYY; in original form ZZZ}

\pubyear{2015}

\newcommand{\Msun}{M_{\odot}}

\newcommand{\be}{\begin{equation}}                                 
\newcommand{\ee}{\end{equation}}                                   
\newcommand{\bea}{\begin{eqnarray}}                                
\newcommand{\eea}{\end{eqnarray}}

\definecolor{gray}{rgb}{.6,.6,.6}                                  %
\definecolor{green}{rgb}{0.6,0,0}                                   %
\definecolor{red}{rgb}{0.6,0,0}                                    

\begin{document}

\label{firstpage}
\pagerange{\pageref{firstpage}--\pageref{lastpage}}
\maketitle

\begin{abstract}
We construct models of static, spherically symmetric shells supported by the radiation flux of a luminous neutron star in the Schwarzschild metric. The atmospheres are disconnected from the star and levitate above its surface. Gas pressure and density inversion appear in the inner region of these atmospheres, which is a purely relativistic phenomenon. We account for the scattering opacity dependence on temperature green by using the Klein-Nishina formula. The relativistic $M_1$ closure scheme for the radiation tensor provides a GR-consistent treatment of the photon flux and radiation tensor anisotropy. In this way we are able to address atmospheres of both large and moderate/low optical depths with the same set of equations. We discuss properties of the levitating atmospheres and find that they may indeed be optically thick, with the distance between star surface and the photosphere expanding as luminosity increases. These results may be relevant for the photosphereric radius expansion X-ray bursts.
\end{abstract}


%%%%%%%%%%%%%%%%%%%%%%%%%%%%%%%%%%%%%%%%%%%%%%%%%%

%
%%%%%%%%%%%%%%%%%%%%%%%%%%%%%%%%%%%%%%%%%%%%%%%%%%%%%%%%%%%%%%%%%%%%%%%%%%
   \section{Introduction}
%%%%%%%%%%%%%%%%%%%%%%%%%%%%%%%%%%%%%%%%%%%%%%%%%%%%%%%%%%%%%%%%%%%%%%%%%%
This paper discusses the structure of spherically symmetric, static,
shell-like atmospheres of extremely luminous, compact, non-rotating stars.
The results are expected to be relevant to the astrophysics of accreting
neutron stars.

Under certain conditions neutron stars may become so luminous that the
forces associated with radiation may exceed the pull of
gravity.  Several systems with super-Eddington luminosity have been
reported \citep{2006csxs.book..157M}, the ``LMC transient'' A0535-668
\citep{1983ARA&A..21...13B} being a particularly clear example.  Super-Eddington luminosities may be achieved in some X-ray bursts
\citep{2006csxs.book..113S}, as well as during accretion of matter in
a semidetached binary, especially in a~ULX \citep{Bachetti2014},
or a detached binary with a Be star as the companion.  At
least in the case of X-ray bursts the radiation field is nearly
spherically symmetric. In most cases, extended periods of time may
occur in which the radiation field and the gas can be taken to be in a quasi-steady state,
i.e., not varying on the dynamical time-scale.  For these reasons we
study the problem of extremely luminous neutron stars under the
simplifying assumptions of steady-state conditions and spherical
symmetry in the Schwarzschild metric.

Contemporary theoretical studies of neutron star atmospheres in X-ray bursts involve sophisticated, spectrally resolved, treatment of the radiation \citep{Suleimanov2011, Suleimanov2012}. However, effects of general relativity (GR) are often neglected for simplicity. In fact, these may be quite important.
The atmospheric structure of luminous stars in GR
has been studied by \citet{Paczynski1986}, who found that the
atmosphere becomes very extended in the Klein-Nishina regime of
scattering opacity. However, in the Thomson regime, as well as in the
Newtonian solutions for either of the scattering regimes, the
atmosphere is geometrically thin. Paczy\'nski and Anderson's results
show that in the case of very luminous neutron stars it would be
inappropriate to expect, and simply speak of, relativistic
``corrections'' to Newtonian solutions. In fact, qualitatively new
results may appear when GR effects are included.

In this paper we report the presence of a~new type of atmospheric
solution for neutron stars radiating at nearly Eddington
luminosities, which is qualitatively different from the ones obtained
in Newtonian physics.  We find that fluid atmospheres of luminous
stars in general relativity may have the form of a shell suspended
above the stellar surface, with the maximum density of the atmosphere
attained on a surface separated from the star by a~``gap'' in which
the atmospheric density and pressure drop precipitously as the stellar
surface is approached.  Such shells may have already been observed in
some X-ray bursts \citep{2011A&A...525A.111I}. A certain group of bursts
indicate radiation-driven ejection of the neutron star gaseous
envelope, e.g., \cite{Tawara1984, White2001, Wolff2005}. Our model is
expected to be of particular interest for modelling these so called
photospheric radius expansion (PRE) bursts 
\citep{Lewin1993}. Unlike most models of atmospheres of
highly luminous sources, e.g., \cite{Kato1994}, the discussed
solutions do not involve dynamical outflows. Our results also differ
from the static extended atmospheres of \citet{Paczynski1986}, which
share with their more familiar Newtonian counterparts the property
that the atmospheric density increases monotonically as the stellar
surface is approached. The key property of our solutions is that neither the density nor the pressure is monotonic as a function of the radius. Analytical models of optically thin Thomson-scattering
shells hovering above the stellar surface of luminous neutron stars were presented by \cite{Wielgus2015}.
In this paper we consistently treat atmospheres of both large and moderate optical depths.

The unusual shell-like structure of our atmospheric solutions can be readily understood as a~consequence of the spatial characteristics of the radiation field and
of gravity. In general relativity, unlike in Newtonian theory, 
the pull of gravity and the radiation flux have a different dependence
on the radial distance from the star. One consequence of this is that
the Eddington luminosity is not a~distance independent 
concept---in fact, typically the flux of radiation has a~stronger dependence on the radius than the effective gravity. Hence,
for a~sufficiently luminous star, the radiation force may balance
gravity only at a~particular radial distance \citep{Phinney1987}.
 Effectively, the Eddington flux is
attained only on a~certain surface  \citep{Bini2009,Oh2010}, 
which is spherical for a~spherically symmetric star.
We refer to this surface as the Eddington Capture
Sphere, or ECS \citep{Stahl2012, Wielgus2012}. 
Inside this surface, radiation force
on an ionized atom exceeds the pull of gravity, outside it gravity
prevails. Thus, the ECS is a~locus of stable
equilibrium positions for test particles \citep{Abramowicz1990,Stahl2012}.

Clearly, an atmosphere may exist,
which is centred on the ECS and thinning out in both directions,
towards and away from the luminous star, with the gas pressure gradient
balancing the difference between the pull of gravity and the radiation
pressure. In the optically thin limit the radiation force is
simply given by the flux of radiation coming from the central star
times the opacity and analytic solutions may be found
  \citep{Wielgus2015}.
Here, we turn our attention to shells of more general optical depth,
which require a~numerical treatment of the gas-radiation interaction.
In our numerical scheme we follow \cite{Levermore1984} and assume
that the radiation tensor is isotropic not in the fluid frame, but in
the ``rest frame'' of the radiation. This leads to the $M_1$ closure.
A~generalization of the $M_1$ scheme to GR has been given in \cite{Sadowski2013}.

For convenience, we parametrize the luminosity of the star by 
the ratio of the luminosity observed at infinity
to the Eddington luminosity,
\be
\lambda=L_\infty/L_{\text{Edd}},
\label{lambda}
\ee
with the standard expression for the latter,
\be
L_{\text{Edd}} = {4 \pi GM m_pc }/{\sigma_{\text T}} = {4 \pi GM c }/{\kappa_{\text T}} \ ,
\ee
for proton mass $m_p$, Thomson cross-section $\sigma_{\text T}$, and Thomson scattering opacity $\kappa_{\text T}$.
In the Schwarzschild metric,
with $|g_{tt}(r)|=1-2GMc^{-2}/r$,
the stellar luminosity at radius $r$ is
\be
L(r)=L_\infty\left(1-\frac{2GM}{rc^2}\right)^{-1},
\label{luminosity}
\ee
and a static balance between gravity and radiation force
with Thomson scattering can only be achieved at one radius
$r=R_{\text{ECS}}$, with
\be
R_{\text{ECS}} \equiv R_{\text{S}}\left[1-(L_\infty/L_{\text{Edd}})^2\right]^{-1} \ ,
\ee
Thus, in terms of the constant introduced in Eq.~(\ref{lambda}),
\be
R_{\text{ECS}}/R_{\text{S}}=1/(1-\lambda^2) \ .
\label{ECS}
\ee
Numerous authors
have shown that test particles initially orbiting the star (at various
radii) will settle on the spherical surface at $r=R_{\text{ECS}}$,
provided that
\be
(1- R_{\text{S}}/R_*)^{1/2}\le\lambda<1 \ ,
\label{range}
\ee
their angular momentum having been removed by radiation drag
\citep{Bini2009,Oh2011,Stahl2012}.
In fact, any point on the ECS is a position of \emph{stable} equilibrium
 in the radial direction 
\citep[and neutral equilibrium in directions tangent
 to the ECS surface,][]{Stahl2012}.

 Unless indicated otherwise, throughout this paper we use geometrical
  units, $G = c = 1$, with $2M$ denoting the Schwarzschild radius
 $R_{\rm {S}}=2GM/c^2$. In the numerical calculations we fix the mass and radius
  of the central spherical star to $M = 1.5 \Msun$, $R_* = 2.5 R_{\rm {S}}$
 (corresponding to $R_* = 11$ km), and consider atmospheres
  consisting of pure ionized hydrogen. All results are given for
  the Schwarzschild spacetime and the signature is assumed to be
  $(-+++)$.

%%%%%%%%%%%%%%%%%%%%%%%%%%%%%%%%%%%%%%%%%%%%%%%%%%%%%%%%%%%%%%%%%%%%%%%%%%
\section{Treatment of radiation}

When treating the radiation as a fluid propagating through
 a possibly optically thick atmosphere, we need to employ a~general formulation
 of the coupled energy-momentum conservation equations for the radiation
 ($R^\mu_{\ \nu}$) and gas ($T^\mu_{\ \nu}$) stress-energy tensors. In relativistic four-notation, the equations take the form
\begin{align}
& (R^\mu_{\ \nu})_{;}\,_{ \mu} = - G_\nu \ , \label{eq:ConsR} \\
& (T^\mu_{\ \nu})_{;}\,_{  \mu} = G_\nu  \ , \label{eq:ConsT}
\end{align}
where $G_\nu$ denotes the radiation four-force density \citep{Mihalas1984},
 a coupling term between gas and radiation. In the
{orthonormal} fluid {rest} frame
 (hereafter denoted with a hat), under the spherical symmetry assumption, the only non-zero components of $G^{\hat{\mu}}$ are $ G^{\hat{t}}$ and $G^{\hat{r}}$, with
\begin{align}
& G^{\hat{t}} = \kappa_{\text{a}} \rho \,(R^{\hat{t} \hat{t}} - 4 \sigma T^4) \ ,
\label{eq:Gt} \\
& G^{\hat{r}} = 
%\left( \kappa_{\text{a}} + \kappa_{\text{s}} \right) \rho R^{\hat{t} \hat{r}}=
  \chi \rho R^{\hat{t} \hat{r}} \ .
\label{eq:RadForceR0}
\end{align}
Here $\chi = \kappa_{\text{a}} + \kappa_{\text{s}} $ denotes the total
opacity coefficient, $\kappa_{\text{a}}$ is the frequency integrated absorption
 opacity, $\kappa_{\text{s}}$ is the scattering opacity, $\sigma$ is the Stefan-Boltzmann constant, and $T$ and $\rho$ are the temperature and rest-mass density of the gas. We neglect the transfer of energy by Compton scattering.

%+++++++++++++++++++++++++++++++++++++++++++++++++++++++++++++++++++++++
\begin{figure*}
\center
\includegraphics[width = 7.0 in, trim = {0mm 0 0 0}, clip]{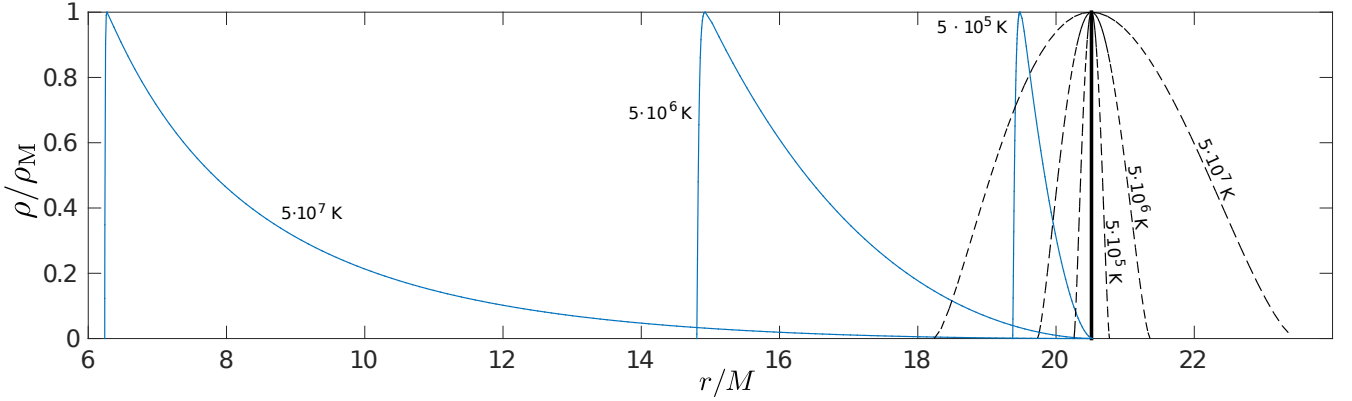} 
\vspace*{-2mm} 
\caption{ Comparison of density profiles for optically thin atmospheres found with Thomson scattering opacity (black, dashed lines) and Klein-Nishina scattering opacity (blue, continuous lines) for luminosity $\lambda = 0.95$. The maximum temperatures are: $T_M = 5 \cdot 10^5$K, $5 \cdot 10^6$ K, $5 \cdot 10^7$ K. Thick black continuous line denotes the test particle ECS location. For the Klein-Nishina atmospheres a significant reduction of the equilibrium radius $R_{\rm E}$ with temperature increase is observed.  }
\label{fig:thinTKN}
\end{figure*}
%+++++++++++++++++++++++++++++++++++++++++++++++++++++++++++++++++++++++

When a static, and spherically symmetric system is to be
considered, only the radial derivatives are of interest.
Eqs.~(\ref{eq:ConsR})-(\ref{eq:ConsT}) then
become ordinary differential equations in the variable $r$.
The ``angular'' components {$R^{({\phi})({\phi})} = R^{({\theta})({\theta})}$}
can be eliminated if one remembers that the radiation tensor
has a~vanishing trace, $R^{\mu}_{\ \mu} = 0$. 
We indicate components in the orthonormal Schwarzschild tetrad
 by indices in parentheses. Under our assumptions this tetrad coincides with the orthonormal fluid rest frame, hence $R^{\hat{\mu} \hat{\nu}} \equiv R^{(\mu)(\nu)}$. This simplifies the calculations greatly, so that Eq. (\ref{eq:ConsR}) yields the following system
\begin{align}
& \frac{1}{r^2}\frac{\text{d}}{\text{d}r} \left( r^2 R^r_{\ t} \right)
 = -G_t\ ,  \label{eq:RtrEvol} \\
& \frac{\text{d}}{\text{d}r} R^r_{\ r}
 = -\frac{ \left(r -{3M} \right) R^t_{\ t}
  + \left(3r- {5M} \right) R^r_{\ r} }{r^2(1-2M/r)}
  - G_r \ . %\label{eq:RrrEvolSys}
\label{eq:RrrEvol}
\end{align}
In general, solving Eq.~(\ref{eq:RrrEvol}) requires knowledge of
the radiative force term $G_r$, as well as the radiation energy density, which
is given by
$R^{(t)(t)}=-R^t_{\ t}$ in the Schwarzschild spacetime.  

We will assume $G^{\hat{t}} = 0 $ throughout this paper. Formally, from Eq.~(\ref{eq:Gt}), this implies that either absorption is negligible or $R^{\hat{t} \hat{t}} - 4 \sigma T^4=0$. The latter corresponds to the condition of local thermodynamic equilibrium (LTE).

Clearly, Eq.~(\ref{eq:RtrEvol}) is decoupled from the system
when $G^{\hat{t}}=0 $. It gives the condition of zero flux divergence, with a~simple solution
\begin{equation}
R^{(t)(r)}(r) = 
R^{tr}(r) = \frac{L_\infty}{4 \pi  r^2(1-2M/r)}  \ .
\label{eq:Rtr2}
\end{equation}
With the radiative flux formula given by Eq. (\ref{eq:Rtr2}),
the $G_r$ component becomes
\begin{equation}
G_r =  g_{rr} e^r_{(r)}G^{(r)}
 =   g_{rr} e^r_{(r)} \chi \rho R^{(t)(r)}
%=  \chi \rho(1-2M/r)^{1/2} R^t_{\ r} \  ,
\label{eq:RadForceR}
\end{equation}
where $ e^r_{(r)} =  |g_{rr}|^{-1/2}$ is a~Schwarzschild tetrad coefficient. To solve Eq.~(\ref{eq:ConsT}) we assume an ideal gas and write the stress energy tensor as
\begin{equation}
T^{\mu}_{\ \nu} = (\rho + p + \epsilon)u^\mu u_\nu + \delta^{\mu}_{\ \nu}p  \ ,
\end{equation}
where $p$ and $\epsilon$ are the pressure and internal energy of the gas
 and $u^\mu$ is its four-velocity. Eq. (\ref{eq:ConsT}) then becomes
\be 
\frac{{ \rm d} p}{{\rm d} r}
 = \frac{-(\rho + p + \epsilon)M}{r^2\left(1-2M/r\right)} + G_r
 = \frac{-(\rho + p + \epsilon)M}{r^2\left(1-2M/r\right)} +
 \frac{ \lambda(\chi/\kappa_T) \rho M}{r^{2}   \left(1-2M/r\right)^{3/2}} \ .
\label{eq:dPressdrEntSys}
\ee
Equation (\ref{eq:dPressdrEntSys}) describes the condition for hydrostatic equilibrium
in the presence of gravitational attraction and a radiation force.
Of course, both pressure and the internal energy of the gas
 contribute to the gravitational attraction in the relativistic framework.

In summary, in our system there are two unknown components, $R^{tt}$ and $R^{rr}$,
 of the radiation stress-energy
tensor, which are related by a single differential equation,
Eq.~(\ref{eq:RrrEvol}), and several gas quantities (pressure, density,...)
also related by one equation, Eq.~(\ref{eq:dPressdrEntSys}). In order to solve
for these quantities, it is necessary to make some additional
assumptions, specifically to adopt an equation of state for the gas and
a closure scheme for the radiation.

In the limit of an isotropic radiation tensor, where
 $\rho_{\rm rad} = -R^t_{\ t} = 3R^r_{\ r} = 3R^\theta_{\ \theta}
 = 3R^\phi_{\ \phi} = 3 p_{\rm rad}$,
% (Eddington closure scheme, cf. subsection \ref{ss:closures}),
 summing  Eqs. (\ref{eq:RrrEvol}) and (\ref{eq:dPressdrEntSys}), we recover
 the familiar equation %{\color{green} eq:IsotropicLimit}
\be 
\frac{{\rm d}  p_{\rm tot}}{{\rm d} r}
 = \frac{-(\rho_{\rm tot} + p_{\rm tot}) M}{r^2\left(1-2M/r\right)}
 \ ,
\label{eq:IsotropicLimit}
\ee
where $p_{\rm tot}$ and $\rho_{\rm tot}$ denote total pressure and total energy density of the gas and radiation mixture, given by
%\begin{equation}
\begin{align}
p_{\rm tot} &= p + p_{\rm rad} , \label{eq:ptot} \\
% = p + (R^r_{\ r} + R^\theta_{\ \theta} + R^\phi_{\ \phi})/3 \ , \label{eq:ptot} \\
\rho_{\rm tot} &= \rho + \epsilon + \rho_{\rm rad} \ .
% = \rho + \epsilon + |R^t_{\ t}| \ .
\label{eq:rhotot}
\end{align}
%\end{equation}
Equation (\ref{eq:IsotropicLimit}) is the correct relativistic hydrostatic balance equation of an \emph{optically thick} gas - radiation mixture.

%
%%%%%%%%%%%%%%%%%%%%%%%%%%%%%%%%%%%%%%%%%%%%%%%%%%%%%%%%%%%%%%%%%%%%%%%%%%
\section{Optically thin polytropic atmospheres}
\label{sub:numthin}
While this paper is mainly concerned with atmospheres of arbitrary
optical depth, we will first briefly discuss the optically thin
solutions. In this regime the radiation stress-energy tensor is known
\emph{a~priori} \citep{Abramowicz1990},
 rendering the radiation transfer description trivial.
This suffices to solve for hydrostatic equilibrium of a polytrope, as
the model reduces to the following simple system of equations
(one ordinary differential equation supplemented by algebraic
ones)
\begin{align}
& \frac{{ \rm d} p}{ { \rm d} r} = \frac{-(\rho + p +
  \epsilon)M}{r^2(1-2M/r)} +
 \frac{\lambda(\kappa_{\rm{s}}/\kappa_{\rm T})\rho M}{r^{2}(1-2M/r)^{3/2}}
 \ ,
\label{eq:dPressdrEnt}\\
& p = K\rho^\Gamma = (\Gamma-1)\epsilon = \rho\frac{k_B T }{\mu m_p}\ ,
\label{eq:algebraThin}
\end{align}
where $K$ is the polytropic constant, the adiabatic index is taken to be
 $\Gamma = 5/3$, the mean molecular mass $\mu=1/2$, and the Thomson scattering opacity $\kappa_{\rm T} = 0.4$ cm$^2$/g (these correspond to pure ionized hydrogen in the non-relativistic limit). We neglect absorption, but account for temperature dependence of the scattering coefficient, corresponding to the (averaged) Klein-Nishina scattering model, i.e., $\chi = \kappa_{\rm {s}} = \kappa_{\rm {KN}}(T)$ . 
The following approximate scattering
opacity 
formula \citep{Buchler76, Paczynski1983, Lewin1993} is
used\footnote{We omit from the denominator of Eq.~(\ref{eq:KNfit})
 an additional factor of
 $\left(1 + 2.7 \cdot 10^{11} {\rho}/{T^2}\right)$, which is appropriate
 in the limit of degenerate matter.},
\be \kappa_{\rm {KN}}(T)
 ={ \kappa_T }
   { \left[ 1 + \left(\frac{T}{4.5 \cdot 10^8 K} \right)^{0.86}\right]^{-1}}
 \ .
\label{eq:KNfit}
\ee
The system of equations (\ref{eq:dPressdrEnt})-(\ref{eq:algebraThin}) can be readily solved numerically.
From Eq.~(\ref{eq:dPressdrEnt}) one finds the location of 
the pressure maximum,
\begin{equation}
R_{\rm{E}} = \frac{2M}{1 -\lambda^2 a_1^2 a_2^2 }\ ,
\label{eq:ECSgeneralized}
\end{equation}
with
\begin{align}
& a_1(T) = {\kappa_{\rm{s}}(T)}/{\kappa_{\rm T}} \ ,
\label{eq:OpacityCorrection1} \\
& a_2(T)
 = \left[1 + \frac{ k_B T }{\mu m_pc^2}\frac{\epsilon + p }{p}\right]^{-1} = \left[1 + \frac{ k_B T_M }{\mu m_pc^2}\frac{\Gamma}{\Gamma-1}\right]^{-1} 
 \ ,
\label{eq:TemperatureCorrection}
\end{align}
where the temperature is taken at its  maximum value, treated as an arbitrary constant parametrizing the family of solutions,
$T= T_M \equiv T(R_{\rm{E}})$.
% As Eq.~(\ref{eq:dPressdrEnt}) differs from Eq.~(\ref{equilibrium}) only by 
Because of 
the relativistic correction
 $(p +\epsilon)$ to density
in  Eq.~(\ref{eq:dPressdrEnt})
and the opacity temperature dependence, the
expression for $R_{\rm{E}}$ differs from that for the test particle $R_{\rm{ECS}} $ by the presence of the correction factors
$a_1,a_2$. 
The first factor gives no correction ($a_1=1$) if the Klein-Nishina modification 
to the Thomson scattering is neglected ($\kappa_{\rm{s}}= \kappa_{\rm T}$).
As long as the temperature is much lower than $1\,$GeV,
or $T_M\ll 10^{13}\,$K,  the second correction, $a_2$, is insignificant,
$|1-a_2|\ll1$. 

For non-relativistic temperatures, i.e., 
when the temperature correction factors are equal to unity, $a_1=1=a_2$,
it is straightforward to find analytic solutions for optically thin atmospheres
\citep{Wielgus2015}. 
In particular, for high luminosities the atmospheric shells are suspended above
surface of the star, with the gas  density falling off on both sides of
the sphere on which it attains its maximum. The results obtained here are qualitatively similar in general, and virtually identical to the analytic solutions in the particular case $a_1=1$.
Figure \ref{fig:thinTKN} compares atmospheres described in this section with the analytic results for polytropic optically thin Thomson scattering atmospheres, discussed in \cite{Wielgus2015}, for the same set of maximum temperatures $T_M$ and fixed luminosity $\lambda = 0.95$. Since the equations are homogeneous in density (and pressure, for a~given temperature), we show density profiles normalized by $\rho_{\rm M}= \rho(R_E)$.

When the luminosity is close to Eddington the atmospheric shells are
suspended much closer to the neutron star for Klein-Nishina solutions
than would be the case for purely Thomson scattering, since the opacity decreases at high temperatures in the Klein-Nishina model. This is because
at high luminosities
the denominator of Eq.~(\ref{eq:ECSgeneralized}) is very small, and even a small
correction to the luminosity parameter significantly changes the value of the denominator, and hence of the
radius  of equilibrium $R_{\rm{E}}$.  

Furthermore, the radial
atmospheric profile of the density becomes extremely asymmetric, the atmosphere
falling off quite sharply towards the star, but being rather extended on 
the side away from the star (i.e., for $r>R_{\rm{E}}$)
 owing to the rapid growth of the running value of $a_1(T)$
as the scattering cross-section increases with decreasing
 temperature,
the gradient of pressure in Eq.~(\ref{eq:dPressdrEnt})
being sensitive---at any  radius $r$---to the difference
 $1-\lambda a_1(T)$.

%
%%%%%%%%%%%%%%%%%%%%%%%%%%%%%%%%%%%%%%%%%%%%%%%%%%%%%%%%%%%%%%%%%%%%%%%%%%
\section{Optically thick LTE atmospheres}
%%%%%%%%%%%%%%%%%%%%%%%%%%%%%%%%%%%%%%%%%%%%%%%%%%%%%%%%%%%%%%%%%%%%%%%%%%
%
Observations show that the atmospheres of radius expansion X-ray bursts
are optically thick, so any model aspiring to address these phenomena needs to allow for  larger optical depths than the ones discussed in the previous section. This, in general, demands solving for the coupled interaction of radiation and gas exchanging energy and momentum through absorption and scattering.

%
%%%%%%%%%%%%%%%%%%%%%%%%%%%%%%%%%%%%%%%%%%%%%%%%%%%%%%%%%%%%%%%%%%%%%%%%%%
\subsection{Closure scheme for the radiation tensor}
\label{ss:closures}
%+++++++++++++++++++++++++++++++++++++++++++++++++++++++++++++++++++++++
%%%%%%%%%%%%%%%%%%%%%%%%%%%%%%%%%%%%%%%%%%%%%%%%%%%%%%%%%%%%%%%%%%%%%%%%%%

When absorption is the only process involved the interaction is
local and solving the radiative transfer equation is
straightforward. However scattering on electrons is often important
 (in the atmospheres of thermonuclear X-ray bursts it is even dominant),
and the intrinsically non-local character of transport in that process
renders Monte Carlo methods ineffective, at the same time necessitating the
computationally expensive use of non-local scattering kernels  in the
radiative transfer equation.

   To solve the equations of gas-radiation interaction and evolution
 one has to make certain assumptions. An effective
   approach is to replace the angle-dependent equation of radiative
   transfer with equations describing evolution of only the first few
   moments of the radiation field. Such an approach, however, requires
   a~\emph{closure scheme}, i.e., extra assumptions for calculating the
   missing components of the radiation stress-energy tensor.  

The simplest approach is Eddington closure, which assumes an isotropic radiation field in
the fluid frame, i.e.,
\begin{equation}
R^{\hat{r} \hat{r}} = R^{\hat{\phi} \hat{\phi}} =R^{\hat{\theta}
  \hat{\theta}} = \frac{1}{3}R^{\hat{t} \hat{t}} \ .
\label{eq:EddingtonClosure}
\end{equation}
In this scheme the complete radiation tensor is determined by a~single
 component, the radiation energy density
$R^{tt}$. However, application of this closure scheme is limited
to the optically thick regime.
A more sophisticated approach
 is afforded by the $M_1$ closure,
 which assumes that the radiation
stress-energy tensor
is isotropic (and the radiative flux vanishes)
in the orthonormal ``rest frame'' of the radiation 
\citep{Levermore1984}. This statement is represented by the
following system of equations \citep{Sadowski2013}, 
\be
R^{\mu \nu} = \frac{4}{3} \bar E u^\mu_R u^\nu_R
 + \frac{1}{3} \bar E g^{\mu \nu} \ ,
\label{eq:M1closure}\\
\ee
where $u^\mu_R$ is the radiation rest frame four-velocity, while $\bar E$ is the radiation energy density in this frame. The system of Eqs.~
(\ref{eq:M1closure}) can be solved uniquely at
any given radius, provided that the radiation energy density $R^{tt}$
and radiation fluxes $R^{t \mu}$ are known (the zeroth and first
moments of the specific intensity). The procedure is to first
calculate $\bar E$ and $u^t_R$, which involves solving two coupled
quadratic equations. The solution is chosen uniquely under the
assumption of $\bar E>0$. The remaining components of $u^\nu_R$ are
then evaluated from the corresponding components of
Eq. (\ref{eq:M1closure}) for $\mu = t$. Finally, the second moments of the specific intensity can be calculated. 

In our case it is enough to find $R^{rr}$ using the
closure scheme,
\begin{equation}
R^{rr} = M_1 \left( R^{tt},R^{tr}\right) \ ,
\end{equation}  
since  under the assumption of spherical symmetry 
 we can then evaluate $R^{\phi \phi}$ and $R^{\theta \theta}$ from the
 zero trace condition on $R^{\mu \nu}$.

%++++++++++++++++++++++++++++++++++++++++++++++++++++++++++++++++++++++++++++++
\begin{figure}
\center
\includegraphics[width = 3.45 in,trim = 0mm 0mm 0mm 10mm, clip]{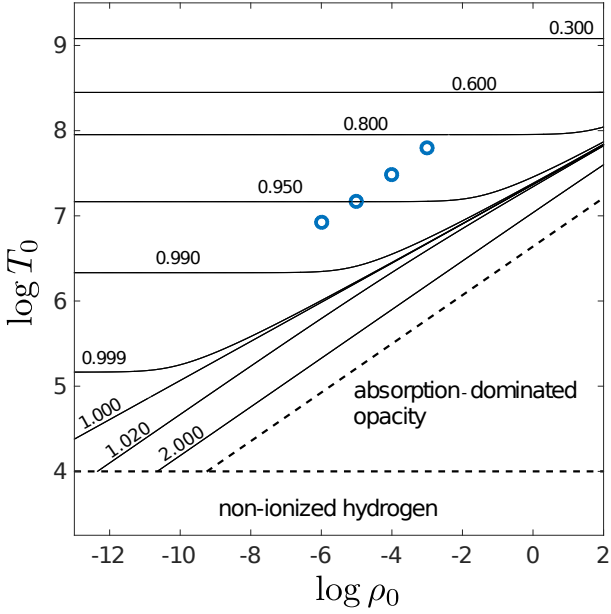} 
\vspace*{-3mm}  
\caption{
 Contour plot of the $a_3$ correction factor, i.e., total opacity to Thomson scattering opacity ratio, Eq.~(\ref{eq:OpacityCorrection3}), as a function of $T_0$ and $\rho_0$. Blue circles correspond to the four optically thick atmosphere solutions with $\lambda = 0.99$, shown in Fig.~\ref{fig:universaldens}. 
 }
\label{fig:parameterSpace}
\end{figure}
%++++++++++++++++++++++++++++++++++++++++++++++++++++++++++++++++++++++++++++++

%++++++++++++++++++++++++++++++++++++++++++++++++++++++++++++++++++++++++++++++
 \begin{figure}
\includegraphics[width = 3.4 in,trim = 0mm 0mm 0mm 0mm]{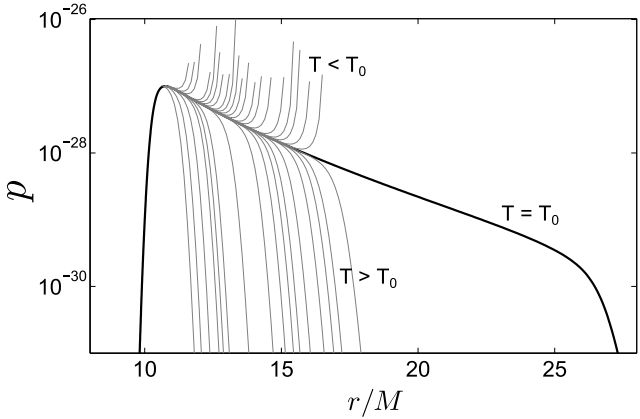}
\vspace*{-5mm}  
 \caption{
 Example of a~levitating atmosphere solution: radial distribution of pressure. Only one initial temperature, for
a given choice of the other parameters, gives a solution which can be extended
to infinity (thick black curve). Other choices of $T_0$ give solutions
with an incorrect temperature at the photosphere (thin curves),
and the pressure going either to infinity, or to zero at a finite radius
$R$, with $R_0<R<\infty$. Pressure in geometrical units.}
\label{fig:Relax}
\end{figure}
%++++++++++++++++++++++++++++++++++++++++++++++++++++++++++++++++++++++++++++++
%%%%%%%%%%%%%%%%%%%%%%%%%%%%%%%%%%%%%%%%%%%%%%%%%%%%%%%%%%%%%%%%%%%%%%%%%%
\subsection{Assumptions of the optically thick model}

In this model we keep the time component of the radiative four-force $G_t$ equal to 
 zero, which is consistent with the LTE assumption,
\begin{equation}
\hat{E} \equiv R^{(t)(t)} = - R^t_{\ t}  = 4 \sigma T^4 \ ,
\label{LTEsys1}
\end{equation}
corresponding to an atmosphere which has had enough time to relax to a~steady state solution. In the opacity model we account for both absorption and scattering,
  $\chi = \kappa_{\rm{a}}+\kappa_{\rm s}$, 
 where $\kappa_{\rm{a}}$ denotes the free-free (bremsstrahlung) opacity given in cgs units by Kramer's formula,
\begin{equation}
\kappa_{\rm{a}} = 6.4 \times 10^{22} T^{-7/2} \rho \ .
\label{eq:brem}
\end{equation}
We also assume a~mean molecular weight of $\mu=1/2$ (pure ionized hydrogen).

Since temperatures of the order of $10^{7} K$ are expected, in the scattering opacity we use the (direction and frequency averaged) Klein-Nishina opacity,
 $\kappa_{\rm s}=\kappa_{\rm{KN}}(T)$, as given by Eq. (\ref{eq:KNfit}). Thus, $\kappa_{s}$ is a~decreasing function of the local
gas temperature and equals the Thomson scattering opacity $\kappa_{\rm T}$
in the low temperature limit.

The system of equations describing our model is then as follows
\begin{align}
& \frac{{\rm d} p}{ {\rm d} r}
 = \frac{-(\rho + p + \epsilon)M}{r^2(1- 2M/r)} + G_r \ ,
\label{eq:LTEdpdr}\\
& \frac{ {\rm d}}{ {\rm d}r} R^r_{\ r}
 = -\frac{ (1 - 3M/r)R^t_{\ t}
 + (3- 5M/r) R^r_{\ r} }{r(1- 2M/r)}
  -G_r \ \label{eq:LTEdRdr} \text{,} \\
& R^{tr}(r) = \frac{L_\infty}{4 \pi  r^2(1-2M/r)} \ ,\\
& G_r =\chi \rho\,{(1-2M/r)^{1/2} R^t_{\ r}} \ \text{,} \\
& p= \frac{k_B}{\mu m_p}\rho T = \frac{2}{3}\epsilon \ ,
\label{eq:ThickEOS} \\
& T = \left( - \frac{ R^t_{\ t}}{ 4\sigma} \right)^{1/4} \ , \\
& R^{rr} = M_1(R^{tt}, R^{tr}) \ , \\
& \chi
 = \kappa_{\rm {a}}(T, \rho) + \kappa_{\rm {KN}}(T) \ .
\label{eq:LTEsys6}
\end{align}
The system of Eqs.~(\ref{eq:LTEdpdr})-(\ref{eq:LTEsys6}) can be solved
 uniquely for the six unknowns, $\rho, p, \epsilon, T, R^t_{\ t},$ and
 $R^r_{\ r}$, as functions of radius.
Equating the right hand side of
Eq. (\ref{eq:LTEdpdr}) to zero we find that the
radius at which the gas pressure attains a~maximum is
\begin{equation}
R_0 = \frac{2M}{1 -\lambda^2 a_3^2 a_2^2}\text{,}
\label{eq:ECSgeneralized2}
\end{equation}
where $a_2$ is a temperature correction similar
to the one present for the polytropic optically thin model, given by
Eq.~(\ref{eq:TemperatureCorrection}) with $T_M$ replaced by $T_0 \equiv T(R_0)$, and
 $a_3$ is a~new correction factor, reflecting the more general
radiation transfer model assumed,
\be a_3 = {\chi(R_0)}/{\kappa_{T}}.
\label{eq:OpacityCorrection3}
\ee

As already remarked,  to high precision $a_2=1$
 at temperatures prevalent in astrophysical
neutron stars.  However, the value of the
$a_3$ parameter has a crucial influence on the position of an
optically thick levitating atmosphere. 
For high temperatures, the opacity
assumes low values because of the Klein-Nishina cross-section reduction
and a much larger flux is required to balance gravity. On the other
hand, dense gas at relatively low temperatures is characterized by
large opacities, because of large absorption. In Fig. \ref{fig:parameterSpace} a~contour plot of the $a_3$ correction factor as a~function of the model parameters is shown. 

For the optically thick models considered here, in general, the radial positions of the density and
pressure peaks do not coincide. The problem is no longer homogeneous in $p$
and does not admit barytropic solutions.
Thus, we need to specify two thermodynamic quantities as boundary values specifying the problem in order to perform the numerical integration. We parametrize
the solutions with $[T_0, \rho_0, \lambda]$, i.e., with the 
temperature at the pressure maximum, $T_0$,
 density at the pressure maximum, $\rho_0 \equiv \rho(R_0)$, and the luminosity
parameter $\lambda$ of Eq.~(\ref{lambda}).
 The initial integration conditions $[p(R_0), R^r_{\ r}(R_0)]$
are calculated from the specified parameters using the ideal gas equation of state (pressure) and the LTE
condition with $M_1$ closure (from $T_0$ we find $R^{tt}$ using the LTE
condition and we evaluate $R^r_{\ r}$ via the $M_1$ closure).
This procedure fixes the values of
 all physical quantities at $R_0$, including the temperature and
opacity corrections $a_2$, $a_3$,
from which $R_0$ itself can be determined,  whereupon the
system of Eqs.~(\ref{eq:LTEdpdr})-(\ref{eq:LTEsys6}) can be integrated in
both directions, starting  at $r=R_0$.
%
%%%%%%%%%%%%%%%%%%%%%%%%%%%%%%%%%%%%%%%%%%%%%%%%%%%%%%%%%%%%%%%%%%%%%%%%%%
\subsection{Outer boundary condition}
\label{sub:bc}
%
%
%
%++++++++++++++++++++++++++++++++++++++++++++++++++++++++++++++++++++++++++++++
 \begin{figure}
 \includegraphics[width = 3.6 in,trim = 0mm 0mm 0mm 4mm]{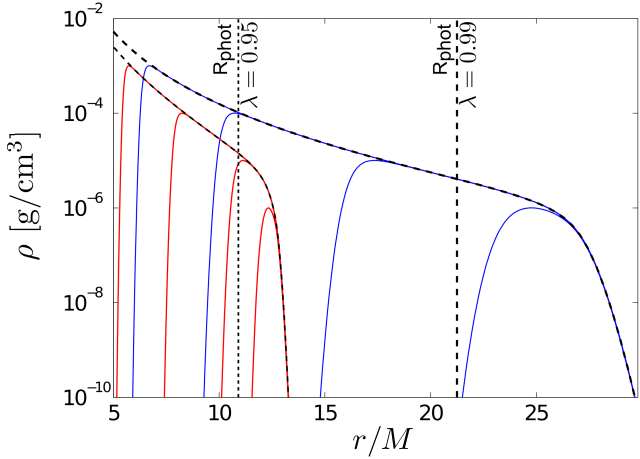}
 \vspace*{-5mm} 
\caption{Comparison of the density profiles for luminosity $\lambda = 0.95$ (family of less extended red curves) and $\lambda = 0.99$ (family of more extended blue curves). The common envelopes of these two families of curves are indicated with dashed black lines. Corresponding locations of photospheres for each family are indicated with the vertical dashed lines. Test particle ECS for such luminosities are located at $20.5M$ and $100.5M$, respectively. This significant reduction of the equilibrium radius is due to the Klein-Nishina effect.}
\label{fig:universaldens}
\end{figure}
%++++++++++++++++++++++++++++++++++++++++++++++++++++++++++++++++++++++++++++++
%
%++++++++++++++++++++++++++++++++++++++++++++++++++++++++++++++++++++++++++++++
 \begin{figure}
 \center
\includegraphics[width = 3.6 in,trim = 0mm 0mm 0mm 7mm]{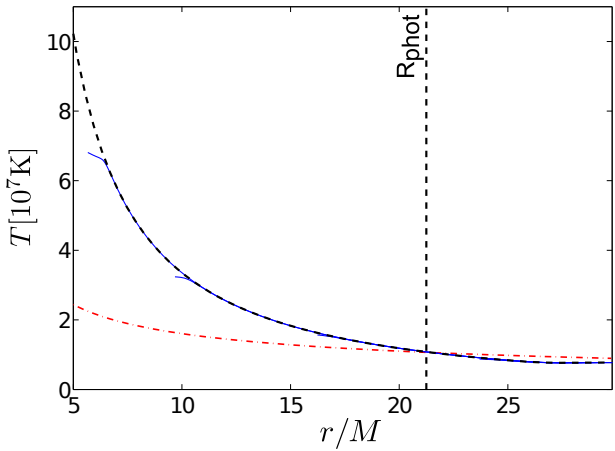}
\vspace*{-5mm} 
 \caption{
 Temperature as a function of radius for LTE atmospheric shells with $\lambda=0.99$. The common envelope indicated with dashed black line, red dash-dotted line corresponds to the effective temperature. Photosphere location indicated with a~vertical dashed line.
}
\label{fig:Temp099}
\end{figure}
%++++++++++++++++++++++++++++++++++++++++++++++++++++++++++++++++++++++++++++++
%
Although the system of Eqs.~(\ref{eq:LTEdpdr})-(\ref{eq:LTEsys6}) can be solved
for any choice of initial parameters $(\rho_0, T_0)$, as described in the previous
subsection, most of the solutions thus obtained would be unphysical, as the
luminosity at infinity would bear no relation to the temperature of the
photosphere. We need to impose an additional condition, that at scattering
optical depth unity, $\tau_{\rm sc}=1$, the LTE temperature is equal to the effective temperature.
Here,
\begin{equation}
\tau_{\rm sc}(r)
 = \int^{\infty}_{r} \rho(r')\, \kappa_{\rm KN}(r')(1-2M/r')^{-1/2}\ \rm{d}r'\ .
\label{BC}
\end{equation}
In practice, it is enough to demand that the pressure is well behaved,
it neither goes to infinity nor vanishes at finite radii. Thus, for any given
initial value of density $\rho_0$, we can reject all trial values of initial
temperature which yield solutions that cannot be extended to infinity.
This determines the value of $T_0$ uniquely for any given $\rho_0$
 (and fixed values of the basic parameters such as $M$ and $\lambda$),
as illustrated in Fig.~\ref{fig:Relax}, where the thick curve corresponds
to the minimum value of $T_0$ among those that yield a finite solution.

To effectively find the correct
solution we use a~numerical relaxation routine assuming $R^{rr}/R^{tr}  = 1$ at the outer boundary, which corresponds to the radiation tensor of a~point source in vacuum. While this condition is an approximation (it is only rigorously fulfilled at infinity), we find that the details of the outer boundary condition have negligible influence on the solution in the region of significant gas density. See the Appendix for some additional comments on the outer boundary condition.

For any (fixed) value of the luminosity, $\lambda < 1$,
we obtain a~family of physical solutions, differing by the density parameter $\rho_0$, related to the total mass of the shell. 
%
%%%%%%%%%%%%%%%%%%%%%%%%%%%%%%%%%%%%%%%%%%%%%%%%%%%%%%%%%%%%%%%%%%%%%%%%%%
\subsection{Properties of the optically thick solutions}
\label{ss:properties}

We find that  for a given luminosity $\lambda$, levitating atmospheres
of optical depth $\tau_{\rm {sc}} > 1$ only exist in a~limited range
of the  $\rho_0$ parameter (density at the pressure maximum). Values
of the density $\rho_0$ that are too low yield optically thin
solutions, values of $\rho_0$ that are too large yield solutions in
which the density decreases monotonically with the radius (the
atmosphere is supported by the surface of the star).

\begin{figure}
\includegraphics[width = 3.3 in,trim = 0mm 0mm 0mm 0mm]{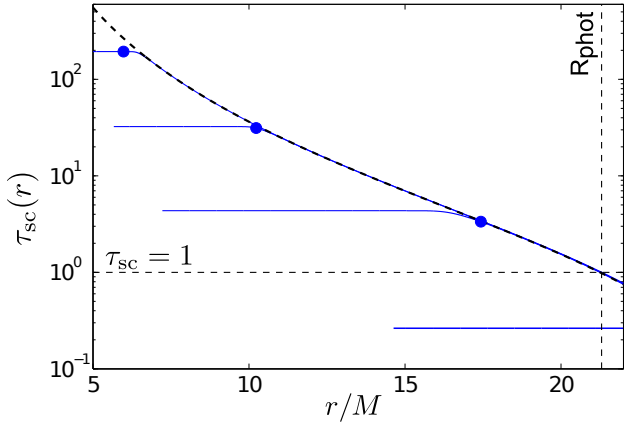}
\vspace*{-5mm} 
 \caption{
Integrated optical depth of the levitating atmospheres obtained for $\lambda = 0.99$. Dots indicate locations of the inner transition surface $R_{\rm tran}$. The vertical dashed line is the common location of the photosphere.
 }  
\label{fig:depth}
\end{figure}

Examples of levitating atmosphere density profiles, found
for $\lambda = 0.95$ and $\lambda = 0.99$, are shown in Fig.~\ref{fig:universaldens}. In the direction towards the star, the atmosphere thins out rather rapidly, so that there is a clear and large gap between
the stellar surface and the levitating atmosphere. Away from the star,
the atmosphere may be quite extended, the thinning out being slower than
exponential, until it becomes optically thin to scattering. In the optically thin outer region the density of the atmosphere decreases
rapidly.
 For a given star (fixed $M, L_\infty$), the base of the levitating
 atmosphere can be located over a~wide range of radii outside the
 star, with its position being determined by the total mass of the
 atmosphere, related to $\rho_0$. In the region of monotonic decrease of $\rho(r)$ all solutions coincide with a common envelope.
The envelope, shown in Figs. \ref{fig:universaldens}-\ref{fig:depth} with thick
black dashed lines, corresponds to the limit of a~monotonic
(non-levitating) solution, with no density inversion (formally $R_0 <R_*$).

Figure \ref{fig:Temp099} illustrates the temperature as a~function of radius for $\lambda=0.99$ solutions. The red dash-dotted line corresponds to an effective temperature, $T_{\rm eff}(r) = [R^{tr}(r)/ \sigma]^{1/4}$. The photosphere location, as found from the back integration of Eq. (\ref{BC}), is consistent with the surface at which the gas temperature is equal to the local effective temperature $T_{\rm eff}(r)$.

\begin{figure}
\includegraphics[width = 3.5 in,trim = 0mm 0mm 0mm 4mm]{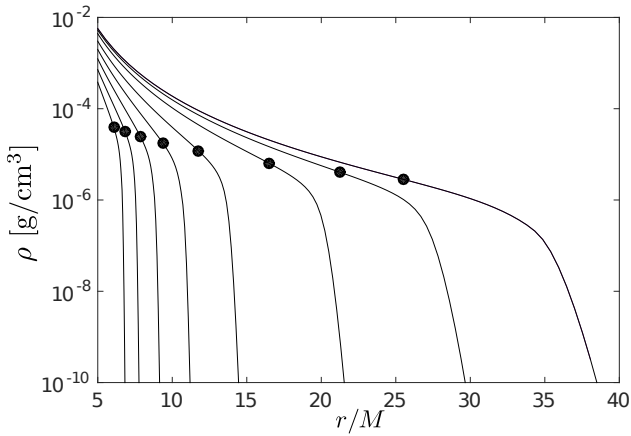}
\vspace*{-5mm} 
 \caption{
 The envelopes of the atmospheric shells
for various values of the luminosity.
From left to right $\lambda= 0.88, 0.90, 0.92, 0.94, 0.96, 0.98, 0.99, 0.995 $.
Black dots indicate the position of the photosphere.}
\label{fig:obwiednie}
\end{figure}

We find that the solutions are strongly dominated by the scattering
opacity, as $\kappa_s/\kappa_a > 10^{5}$ for all radii. Radiation
pressure strongly dominates over the gas pressure, with $p/p_{\rm rad}
< 10^{-3}$ throughout the domain. The solutions vary from optically thin to scattering optical depths of the order of $10^3$. Optical depths of solutions for $\lambda = 0.99$, calculated according to Eq. (\ref{BC}), are shown in Fig. \ref{fig:depth}. An obvious property of the levitating atmospheres is that there exists also an optically thin region in the inner part of the shell. It is located below a~transition surface $r = R_{\rm tran}$, where the optical depth integrated \emph{from the inside} is equal to unity, $\tau_{\rm in}(R_{\rm tran}) = \tau_{\rm sc}(R_*) - \tau_{\rm sc}(R_{\rm tran}) = 1$. These locations are indicated in Fig. \ref{fig:depth} with blue dots.\\
%++++++++++++++++++++++++++++++++++++++++++++++++++++++++++++++++++++++++++++++
%++++++++++++++++++++++++++++++++++++++++++++++++++++++++++++++++++++++++++++++
%
The set of universal density profile envelopes, parametrized with
$\lambda$, is shown in Fig. \ref{fig:obwiednie}. Since the properties
of these envelopes do not depend on $\rho_0$, the location of the
photosphere is common to all the optically thick solutions at fixed
$M, L_\infty$ and tends to larger radii as luminosity
increases, cf. Fig. \ref{fig:photosphere}.
%
%++++++++++++++++++++++++++++++++++++++++++++++++++++++++++++++++++++++++++++++
%++++++++++++++++++++++++++++++++++++++++++++++++++++++++++++++++++++++++++++++

%****************************************************************************
\subsection{Stability of LTE solutions}
%****************************************************************************

Convective stability of an optically thick relativistic atmosphere is determined by the Schwarzschild stability criterion \citep{Thorne1966},
\begin{equation}
S(r) = \frac{1}{\Gamma} \frac{ {\rm d} \log p_{\rm tot}}{{ \rm d} r} - \frac{1}{\rho_{\rm tot} + p_{\rm tot}} \frac{\rm  d \rho_{\rm tot} }{ {\rm  d} r} > 0 \ ,
\label{eq:SchwarzCrit}
\end{equation}
where the total pressure and density, $p_{\rm tot}$ and $\rho_{\rm tot}$, are calculated according to Eqs. (\ref{eq:ptot})-(\ref{eq:rhotot}). The condition (\ref{eq:SchwarzCrit}) can be readily obtained by linearly perturbing our simplified equation (\ref{eq:IsotropicLimit}).
We use $\Gamma = 4/3$, since our atmospheres are strongly radiation
pressure dominated ($p/p_{\rm rad} < 10^{-3}$). 

\begin{figure}
\center
\includegraphics[width = 3.3 in,trim = 0mm 0mm 0mm 0mm]{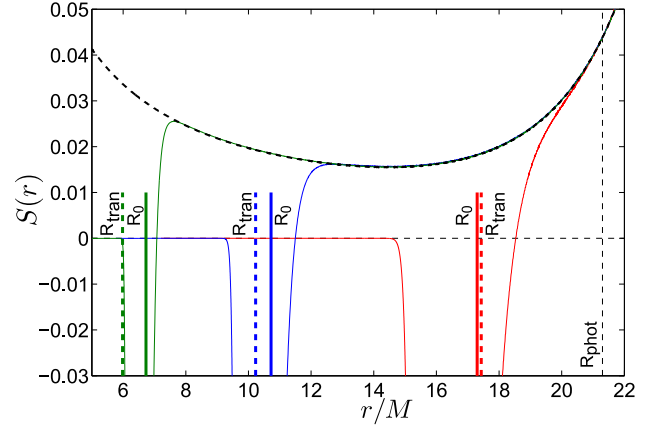}
\vspace*{-5mm} 
\caption{Schwarzschild stability criterion for the levitating atmospheres of $\lambda = 0.99$. The negative values indicate possible convective instability. Curves for the density parameter $\rho_0 = 10^{-3}, 10^{-4}, 10^{-5} \text{g/cm$^3$}$ (left to right).}
\label{fig:SchwStab}
\end{figure}

In Fig. \ref{fig:SchwStab} we show the radial distribution of $S(r)$ for
levitating atmospheres calculated for 
$\lambda = 0.99$, $\rho_0 =10^{-3}, 10^{-4}, 10^{-5} \text{g/cm$^3$}$.
 Positive values of $S(r)$
correspond to convective stability of the outer region of the atmospheres. The
limiting envelope of levitating atmospheric solutions (dashed black line) has $S(r) > 0$ everywhere. Figure \ref{fig:SchwStab} indicates marginal stability, $S(r) = 0$, in the region between the stellar surface and the inner edge of the levitating atmosphere, suspended above the star. This is in agreement with the analytic limit of $S(r)$ for pure radiation and zero gas density.

The situation in the inner region of the atmospheres is somewhat more complicated. While Fig. \ref{fig:SchwStab} formally indicates convective instability, $S(r)<0$, near the transition radius $r = R_{\rm tran}$ the large optical depth condition is not met and hence the criterion itself is not strictly valid. Moreover, near-Eddington radiation flux is expected to have a~strong stabilizing influence, damping the motion of the optically thin fluid,  \cite{Stahl2013}, and thus hindering the development of instabilities. 
For instance, Abarca \& Klu\'{z}niak (in prep.) find that the fundamental radial mode of atmospheric oscillations is overdamped. 

%++++++++++++++++++++++++++++++++++++++++++++++++++++++++++++++++++++++++++++++
%++++++++++++++++++++++++++++++++++++++++++++++++++++++++++++++++++++++++++++++

There clearly exists a necessity for a~more general convective
stability criterion, that would remain valid regardless of the optical
depth. This, however, is beyond the scope of this work and will
be a subject of future investigations.

%****************************************************************************
\section{Discussion: photospheric radius expansion bursts}
%****************************************************************************

The main result of this paper is that the atmospheres of luminous
neutron stars may form static shells, suspended above the neutron star
surface by the force of radiation. These shells may be optically thick
or thin, depending on the amount of matter forming the shell,
presumably ejected from the neutron surface in a~luminous burst of
thermonuclear origin. 

Even if enough matter is ejected to initially form an optically thick
atmosphere, it may easily become optically thin as luminosity
increases and the envelope expands (if an envelope of fixed mass $M_e$ is expanding, its optical depth goes down approximately with the inverse square of radius). This further justifies the necessity of a~model capable of addressing properly the regime of optically thick gas and possible transition to the optically thin regime. An interesting feature of our model is that if the photosphere is formed, its  properties (location, temperature) only depend on the luminosity parameter $\lambda$ and not on the mass of the envelope.

A certain group of X-ray bursts exhibit particularly strong peak luminosities, approaching the Eddington limit \citep{Lewin1993, 2006csxs.book..113S}.~A~strong radiative force may push the gaseous envelope of the neutron star away from the stellar surface. 
In some luminous bursts it is observed that the emitting surface, inferred from the effective temperature and the luminosity, increases during the early stages of the bursts and then decays to its initial value during the so called touch-down phase. This group of bursts is referred to as PRE bursts. It was argued that the burst luminosity almost exactly reaches the Eddington luminosity during the expansion phase, \citep{Kato1994}. The main argument to support this strong claim is that any excess energy from super-Eddington flux would be efficiently converted to kinetic energy, resulting in dynamical outflows. Sub-Eddington flux could not, on the other hand, explain the photospheric radius expansion, inferred from the observed spectra.

\begin{figure}
\includegraphics[width = 3.28 in,trim = 0mm 0mm 0mm 0mm]{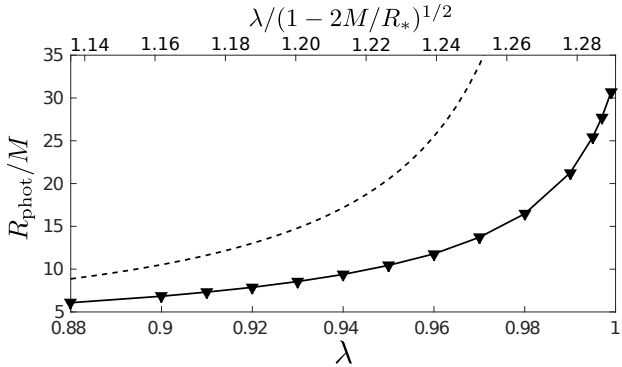}
\vspace*{-2mm} 
 \caption{ 
 Photosperic radius, $R_ {\rm phot}$ and the test particle Eddington capture sphere radius $R_{\rm ECS}$ (dashed lines).}
\label{fig:photosphere}
\end{figure}

\begin{figure}
\includegraphics[width = 3.28 in,trim = 0mm 0mm 0mm 0mm]{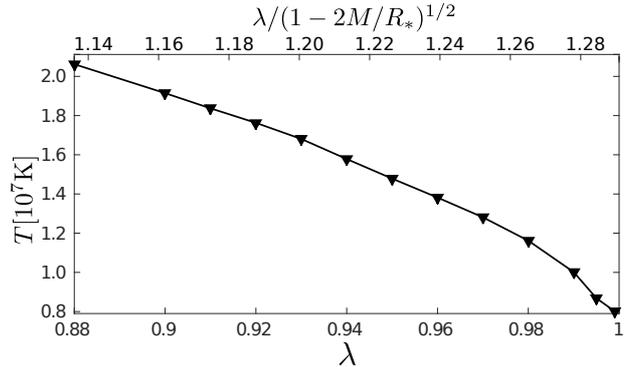}
\vspace*{-2mm} 
 \caption{
 Photospheric temperature as a function of luminosity $\lambda$.}
\label{fig:photosphereT}
\end{figure}

Observations of PRE bursts can reveal important knowledge about the
dependence between neutron star mass and radius \citep{Damen1990,
  Ozel2006, Ozel2015} and it is reasonable to expect that better understanding
of the relations between luminosity, photosphere location and
temperature in PRE bursts should provide more detailed insight. For
instance, in the basic models, the location of the photosphere is assumed
\citep{Damen1990} to coincide with what was later recognized to be the test-particle
Eddington capture sphere \citep{Stahl2012}. 
We find that this is a~significant overestimation, since
the photospheric radius of the levitating atmospheres is typically situated much closer to the stellar surface than the test particle ECS,
(Fig. \ref{fig:photosphere}). We also observe that it is not necessary
for the flux to be of Eddington value for the photosphere to start expanding. Figure \ref{fig:photosphere} indicates that the expansion
begins at luminosity about $0.85 L_{\rm Edd}$ and progresses with the
increase of luminosity. For luminosity equal to $0.99 L_{\rm Edd}$ the
photospheric radius expands by a factor of about 4, while simultaneously cooling down by a~factor of 2, Fig. \ref{fig:photosphereT}. This seems to be in agreement with a~typical expansion magnitude inferred from the observational data, see, e.g., \cite{Zand2013}. In Figs. \ref{fig:photosphere} and \ref{fig:photosphereT} the horizontal axes are labelled with $\lambda$ as used throughout this paper, as well as and with luminosity normalized by the factor of $L_{\rm Edd}(1 - 2M/R_*)^{1/2}$, corresponding to the equilibrium luminosity at the stellar surface $r = R_*$. We note that the discussed luminosities are mildly super-Eddington if the latter is adopted as an "Eddington luminosity" unit.

In conclusion, we suggest that detailed modelling of near-Eddington photospheric expansion bursts should take into account the effects described in this work.

%
%++++++++++++++++++++++++++++++++++++++++++++++++++++++++++++++++++++++++++++++
%++++++++++++++++++++++++++++++++++++++++++++++++++++++++++++++++++++++++++++++
%++++++++++++++++++++++++++++++++++++++++++++++++++++++++++++++++++++++++++++++
%++++++++++++++++++++++++++++++++++++++++++++++++++++++++++++++++++++++++++++++

\section*{Acknowledgements}
We thank Omer Blaes, Feryal \"Ozel, and David Abarca for interesting comments on the subject of this paper. This research was partly supported by Polish NCN grants
UMO-2011/01/B/ST9/05439 and UMO-2013/08/A/ST9/00795
 as well as the Czech ASCRM100031242 CZ.1.07/2.3.00/20.0071 ``Synergy'' (Opava) project. MW also acknowledges the support of the Foundation for Polish Science within the START programme.

\bibliography{Atm2}
\bibliographystyle{mnras}

%\newpage
\appendix

\section{Comments on the $M_1$ closure}

\begin{figure}
\center
\includegraphics[width = 3.4 in, trim= {0mm 0 0 15}, clip]{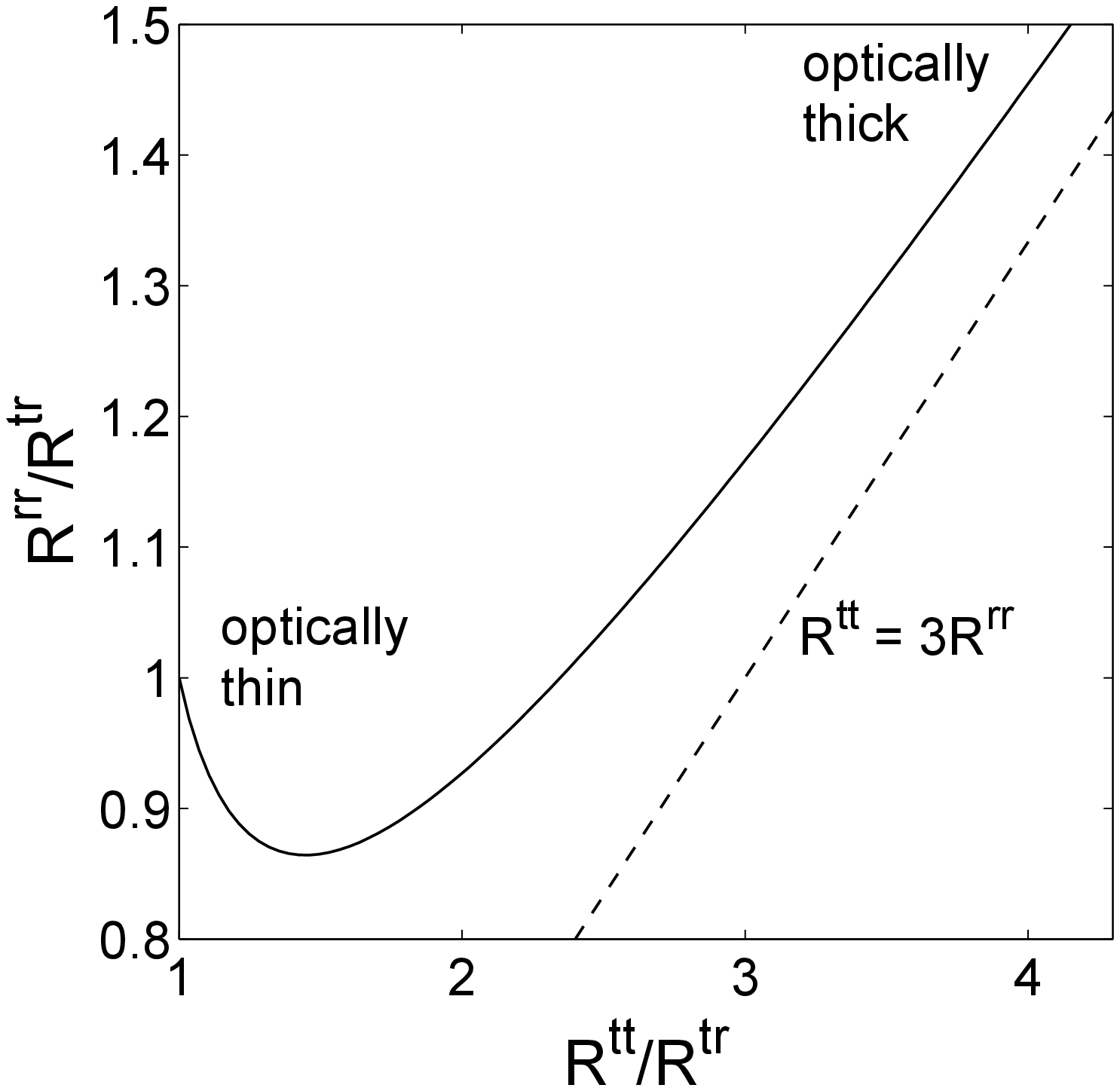}  
\vspace*{-2mm} 
\caption{%{\color{green}fig:M1plot}
The ratio $R^{rr}/R^{tr}$
 versus $R^{tt}/R^{tr}$ calculated with
 the $M_1$ scheme for $M/r \rightarrow 0$. 
Notice the non-injective character of the $M_1$ 
closure scheme in the optically thin regime.}
\label{fig:M1plot}
\end{figure}

In the limit of $M/r \rightarrow 0$ an analytic expression for the $M_1$ closure can be given
\begin{equation}
\frac{R^{rr}}{R^{tr}} = \frac{2 - a^2 + a(4a^2 -3)^{1/2}} {a + (4a^2-3)^{1/2}} \ ,
\label{eq:analyticM1}
\end{equation}
where $a = R^{tt}/R^{tr}$. The corresponding curve is shown in Fig.~\ref{fig:M1plot}. With Eq.~(\ref{eq:LTEdRdr}) being a differential equation for the $R^{rr}$ component, in this work we are actually interested in the ``inverse $M_1$'' problem,
i.e., in finding the $R^{tt}$ component, given $R^{rr}$ and $R^{tr}$ at
every step of the numerical integration of Eq.~(\ref{eq:LTEdRdr}). It
is worth noticing that while the $M_1$ closure scheme is
a~unique procedure, it is not an injective function of
$(R^{tt}, R^{tr})$, meaning that a~given $R^{rr}$ may correspond to more
than one pair $R^{tt}/R^{tr}$, see
Fig.~\ref{fig:M1plot}. Closer inspection of the formula (\ref{eq:analyticM1}) reveals that the minimum of the $M_1$ curve corresponds to $\beta = u^r_R / u^t_R = c/\sqrt{3}$ and separates the right ``subsonic photon gas'' optically thick branch from the left ``supersonic photon gas'' optically thin branch. The numerical relaxation procedure that we utilize is necessary for the solution to pass through that ``sonic point'', and allow for a~continuous transition from the optically thick regime of radiation trapped in the gas to the optically thin regime of freely streaming photons, forced by the outer boundary condition. 

We note that in vacuum for an isotropic radiation
stress tensor, i.e., for $G_r=0$ and $R^t_{\ t}=-3R^r_{\ r}$,
 Eq.~(\ref{eq:RrrEvol}) takes the form
\be
\frac{\text{d}}{\text{d}r} R^r_{\ r}
 = -\frac{ 4M}{r^2(1-2M/r)}\, R^r_{\ r} 
  \text{,}
\ee
with the solution $R^r_{\ r}=p_0/(1-2M/r)^2$, where $p_0$ 
is an integration constant of dimension pressure.
The energy density of this isotropic radiation field scales correctly
with the fourth power of the redshift factor $(1+z)$: 
\be
R^{(t)(t)}=3R^{(r)(r)}={3p_0}{\left(1-\frac{2M}{r}\right)^{-2}}.
\label{eq:nogas}
\ee
We find that in the region between the stellar surface and the atmosphere
our numerical solutions follow Eq.~\ref{eq:nogas} quite closely.
%
%+++++++++++++++++++++++++++++++++++++++++++++++++++++++++++++++++++++++
%+++++++++++++++++++++++++++++++++++++++++++++++++++++++++++++++++++++++

% Don't change these lines
\bsp	% typesetting comment
\label{lastpage}
\end{document}